\documentstyle[12pt,aps,floats,amsmath]{revtex}
\input BoxedEPS
\SetRokickiEPSFSpecial  
\HideDisplacementBoxes

\tighten

\newcommand{\Cite}[1]{[{\tt #1}]\cite{#1}\relax}
\let\Cite\cite

\newcommand{\numeq}[2]{\begin{equation}
#2
\label{#1}
\end{equation}}

\newcommand{\bra}[1]{\bigl\langle #1 \bigr| }
\newcommand{\ket}[1]{\bigl| #1 \bigr\rangle}
\newcommand{\q}{\quad}

\begin{document}
\draft
\title{Extended Gari-Kr\"umpelmann model fits to nucleon 
electromagnetic form factors}


\author{Earle L. Lomon}
\address{Center for Theoretical Physics\\
Laboratory for Nuclear Science and 
Department of Physics\\ Massachusetts Institute of Technology, Cambridge, 
Massachusetts 02139\\
\rm MIT-CTP-3111}
\maketitle
\begin{abstract}
Nucleon electromagnetic form factor data (including recent data) is fitted with
models that respect the confinement and asymptotic freedom properties of QCD. 
Gari-Kr\"umpelmann (GK) type models, which include the major vector meson pole
contributions and at  high momentum transfer conform to the predictions of
perturbative QCD, are combined  with H\"ohler-Pietarinen (HP) models, which also
include the width of the $\rho$ meson and the addition of higher mass vector
meson exchanges, but do not evolve into the explicit form of PQCD at high
momentum transfer.  Different parameterizations of the GK model's hadronic form
factors, the effect of including  the width of the $\rho$ meson and the addition of
the next (in mass) isospin~1  vector meson are considered.  The quality of fit and
the consistency of the parameters select three of the combined HP/GK type
models.  Projections are made to the higher  momentum transfers which are
relevant to electron-deuteron experiments. The projections vary little for the
preferred models, removing much of the ambiguity in electron-nucleus scattering
predictions.
\end{abstract}



\section{INTRODUCTION}

	The electromagnetic form factors (emff) of the neutron and proton contain 
all the information about the charge and current distribution of these
baryons,  providing strong constraints on the fundamental theory of strong
interactions.   In addition the predictions for the emff of many-nucleon
systems are sensitive  to the input nucleon emff, as well as the many-body
effects one would like to  determine.  For the first aspect one would like to
have an accurate description  of the data in a form closely linked to the
fundamental theory.  For the second it is convenient to have a simple
analytic form to embed in the many-body  calculation.  In the past models
with different physical constraints, equally well fitted to the available data, have
predicted nucleon emff which differ sufficiently to induce large ambiguities in 
deuteron and heavier nucleus emff predictions.  This has limited what can be
learned about nuclear forces and meson-exhange current effects.  The analysis
here shows that, by combining the important physical features of past models
and the data set now available, the few models which fit the data well, and with
parameters most consistent with other reactions, produce small variations in the
nucleon predicted emff over an extended range.

	Accepting QCD as the fundamental theory of strong interactions, the emff 
can be described by perturbative QCD (PQCD) at very high momentum
transfers.  At  low momentum transfers the confinement property of QCD
implies an effective  hadronic description with vector meson dominance
(VMD, the coupling of the  photon to vector mesons that couple in turn to
the nucleons).  Early models of  the nucleon emff were based on VMD alone
\Cite{IJL,HO} including the $\rho$, 
$\omega$, and $\phi$ poles and the cut associated with the $\rho$ width, but
with  several phenomenological higher mass poles added. Gari and
Kr\"umpelmann~\Cite{GK} restricted the VMD contribution to the 
$\rho$, $\omega$, and  $\phi$  poles,  but added factors and terms which explicitly
constrained the asymptotic momentum  transfer behavior to the scaling behavior
of PQCD.  The additional factors,  specified in the next section, are, in
effect, hadronic form factors. We fit a series of four GK type models (varying only in
the details of the hadronic form  factors, as motivated in Section~\ref{s:2}) to the
present data set.  In addition to the GK type models we consider a group of models
(generically designated DR-GK) that use the analytic approximation of
\Cite{MMD} to the dispersion integral  approximation for the $\rho$ meson
contribution (similar to that  of HP \Cite{HO}), modified by the hadronic form  factors of
the type we use with the GK model, and the addition of the $\rho'$ (1450) pole. 
These additions result in a better fit to the data than we obtain  with only the GK model
\Cite{GK} and minor variants of the hadronic form factors.

In this paper we fit the world data set for $G_{Ep}$, $G_{Mp}$, $G_{En}$, $G_{Mn}$
and  
$R_p =\mu_p G_{Ep}/G_{Mp}$. The last quantity, $R_p$, is a direct result of a
recent  measurement~\Cite{MKJ} with a polarized electron beam.  We find similar 
results with the GK type models for three different parameterizations of the 
hadronic form factors (the fourth fits poorly), all three of the fits being reasonable
when the inconsistency  of the of the data, particularly for the neutron, is taken into
account (some  of the data sets must have large systematic errors, unless the emff
oscillate  over unnaturally small momentum transfer scales). With the extended
DR-GK type models  described above, qualitatively better fits are obtained for all
four  parameterizations of the hadronic form factors.

	In Section~\ref{s:2} we will specify the models and parameters.  Section~\ref{s:3}
will  summarize the data set and the optimization procedure, while Section~\ref{s:4}
will  present the results in comparison with each other and the original GK fit.  We 
extrapolate beyond the present experimental range of momentum transfer where
necessary for predicting available deuteron emff, and comment on the differences
between  the models in the extended range.  For the three models with the lowest
(nearly equal)  $\chi^2$ fits to the data and parameters most consistent with other
reactions, the differences are small.  These DR-GK models are consistent with the
requirements of dispersion relations and of QCD at low and high momentum
transfer. 

If these models are used as
input in many-nucleon form factor calculations  they will provide approximately stable
results consistent with the nucleon data.   Discrepancies with the many-nucleon data
can then be attributed to deficiencies  in the many-body wave function,
meson exchange currents or relativistic  corrections.

\section{Nucleon emff models}\label{s:2}

	The emff of a nucleon are defined by the matrix elements of the 
electromagnetic current~$J_\mu$
\numeq{e1}{
\bra{N(p')}\, J_\mu \,\ket{N(p)} = e \bar u (p') \Bigl\{
\gamma_\mu F^N_1 (Q^2) + \frac i{2m_N} \sigma_{\mu\nu} Q^\nu F^N_2(Q^2)
\Bigr\} u(p) 
}
where $N$ is the neutron, $n$, or proton, $p$, and $-Q^2= (p'-p)^2$ is the square  of
the invariant momentum transfer.  $F_1^N(Q^2)$ and $F_2^N(Q^2)$ are respectively 
the  Dirac and Pauli form factors, normalized at $Q^2=0$ as
\numeq{e2}{
F^p_1(0) = 1,\q F^n_1(0)=0,\q F^p_2(0) = \kappa_p,\q F^n_2(0)= \kappa_n\ .
}
Expressed in terms of the isoscalar and isovector electromagnetic currents
\numeq{e3}{
2F^p_i = F^{\null{is}}_i + F^{\null{iv}}_i, \q
2F^n_i = F^{\null{is}}_i - F^{\null{iv}}_i, \q
(i=1,2)\ .
}
The Sachs form factors, most directly obtained from experiment, are then
\begin{eqnarray}
G_{\mathrm{EN}} (Q^2) &=& F^N_1 (Q^2) -\tau F^N_2(Q^2)\nonumber\\
G_{\mathrm{MN}} (Q^2) &=& F^N_1 (Q^2) +  F^N_2(Q^2),\q \tau =
\frac{Q^2}{4m_N}\ .
\label{e4}
\end{eqnarray}
The model of Gari and Kr\"umpelmann \Cite{GK} prescribes the 
following form for the four emff:
\begin{eqnarray}
F^{\null{iv}}_1(Q^2) &=&  \frac{g_\rho}{f_\rho}
\frac{m^2_\rho}{m^2_\rho + Q^2} F^\rho_1(Q^2) + \Bigl(
1-\frac{g_\rho}{f_\rho}\Bigr)
F^{\null D}_1(Q^2)\nonumber\\
F^{\null{iv}}_2(Q^2) &=&  \kappa_\rho\frac{g_\rho}{f_\rho}
\frac{m^2_\rho}{m^2_\rho + Q^2} F^\rho_2(Q^2) + \Bigl(
\kappa_{\null v} -\kappa_\rho\frac{g_\rho}{f_\rho}\Bigr)
F^{\null D}_2(Q^2)\nonumber\\
F^{\null{is}}_1(Q^2) &=&  \frac{g_\omega}{f_\omega}
\frac{m^2_\omega}{m^2_\omega + Q^2} F^\omega_1(Q^2) + 
\frac{g_\phi}{f_\phi}\frac{m^2_\phi}{m^2_\phi + Q^2} F^\phi_1(Q^2) +
\Bigl(
1-\frac{g_\omega}{f_\omega}\Bigr)
F^{\null D}_1(Q^2)\label{e5}\\
F^{\null{is}}_2(Q^2) &=& \kappa_\omega \frac{g_\omega}{f_\omega}
\frac{m^2_\omega}{m^2_\omega + Q^2} F^\omega_2(Q^2) + 
\kappa_\phi \frac{g_\phi}{f_\phi}\frac{m^2_\phi}{m^2_\phi + Q^2} F^\phi_2(Q^2) +
\Bigl(
\kappa_{\null s} - \kappa_\omega \frac{g_\omega}{f_\omega} -
\kappa_\phi\frac{g_\phi}{f_\phi}\Bigr) F^{\null D}_2(Q^2)
\nonumber
\end{eqnarray}
where the pole terms are those of the $\rho$, $\omega$, and  $\phi$  mesons, and
the  final term of each equation is determined by the asymptotic properties of
PQCD.   The $F_i^\alpha$, $\alpha = \rho$,  $\omega$, or $\phi$ are the
meson-nucleon form factors,  while the $F_i^D$ are effectively quark-nucleon form
factors.

	In the final form used by GK, called Model 3 in~\Cite{GK}, the above  hadronic
form factors are parameterized in the following way:
\begin{eqnarray}
F^{\alpha,D}_1(Q^2) &=&  \frac{\Lambda^2_{1,D}}{\Lambda^2_{1,D} +\tilde Q^2}
\frac{\Lambda^2_2}{\Lambda^2_2 +\tilde Q^2}\nonumber\\
F^{\alpha,D}_2(Q^2) &=&   \Bigl(\frac{\Lambda^2_{1,D}}{\Lambda^2_{1,D} +\tilde
Q^2}\Bigr)^2
\frac{\Lambda^2_2}{\Lambda^2_2 +\tilde Q^2}\nonumber\\
F^\phi_1(Q^2) &=& F^\alpha_1\Bigl(\frac{Q^2}{\Lambda^2_1 + Q^2}
\Bigr)^{1.5}\ , \quad F^\phi_1(0) = 0\nonumber\\
F^\phi_2(Q^2) &=&
F^\alpha_2\Bigl(\frac{\Lambda^2_1}{\mu^2_\phi}\frac{Q^2+\mu^2_\phi}{\Lambda^2_1
+ Q^2}
\Bigr)^{1.5}
\label{e6} \\
\mbox{with } \tilde Q^2 &=& Q^2\frac{\ln\bigl[(\Lambda^2_2 +
Q^2)/\Lambda^2_{\mathrm{QCD}}\bigr]}{\ln(\Lambda^2_2/\Lambda^2_{\mathrm{QCD}})}\
, \quad\mbox{where
$\alpha=\rho,\omega$.} \nonumber
\end{eqnarray}

This parameterization, together with Eq.~(\ref{e5}), guarantees that the 
normalization conditions of Eq.~(\ref{e2}) are met and that asymptotically
\begin{align}
F^i_1 &\sim \bigl[ Q^2 \ln(Q^2/\Lambda^2_{\mathrm{QCD}})\bigr]^{-2}\nonumber\\
F^i_2 &\sim F^i_1/Q^2 \label{e7}\\
i &= is, iv \nonumber
\end{align}
as required by PQCD.  When fitted to the data set described in section III,
the  result is  here called Model~GK(3).

	In their Model 1 (fitted only to the proton data) GK associated the  helicity
flip hadronic form factors, $F_2$, with the quark-gluon scale cut-off 
$\Lambda_2$.  However in model 3, in fitting to the available data, they chose
to  associate the helicity flip with the meson scale cut-off $\Lambda_1$, as 
incorporated in Eq.~(\ref{e6}).  To investigate the effect of this change we also fit our 
data set, in Model~GK(1), with the hadronic form factors of GK Model~1, for which
\begin{equation}
F^{\alpha,D}_2 = \frac{\Lambda^2_{1,D}}{\Lambda^2_{1,D} +\tilde Q^2}
\Bigl(\frac{\Lambda^2_2}{\Lambda^2_2 +\tilde Q^2}\Bigr)^2
\label{e8} 
\end{equation}
replaces the expressions in Eq.~(\ref{e6}).

	In both of the above parameterizations the logarithmic $Q^2$ dependence of 
PQCD is approached through a form factor determined by the $\Lambda_2$ and 
$\Lambda_{\mathrm{QCD}}$ cut-offs.  In our Model~GK$'$(1) we replace
$\Lambda_2$ with
$\Lambda_D$ for  that factor which relates to the quark-nucleon vertex:
\begin{equation}
\tilde Q^2  =  Q^2\frac{\ln\bigl[(\Lambda^2_D +
Q^2)/\Lambda^2_{\mathrm{QCD}}\bigr]}
{\ln(\Lambda^2_D/\Lambda^2_{\mathrm{QCD}})}\ .
\label{e9} 
\end{equation}
Otherwise Model~GK$'$(1) is the same form as Model~GK(1). A similar replacement 
was attempted for Model~GK(3), but the best fit was substantially worse with the 
modification.

The next group of models replaces the $\rho$ meson pole terms in $F^{(iv)}_1$  and
$F^{(iv)}_2$ (Eq.~5)with the well established $\rho'$ (1450) meson pole term, and adds the
$\rho$ meson term from the dispersion relation in approximate analytic form~\Cite{MMD}:
\begin{align}
F^{\null{iv}}_1(Q^2) &= N
\frac{1.0317 + 0.0875(1+Q^2/0.3176)^{-2}}{(1+Q^2/0.5496)}
   F^\rho_1(Q^2)\nonumber\\
 &\qquad{}  + \frac{g_{\rho'}}{f_{\rho'}}
\frac{m^2_{\rho'}}{m^2_{\rho'} + Q^2} F^\rho_1(Q^2)
+ \Bigl( 1-1.1192\, N -  \frac{g_{\rho'}}{f_{\rho'}} \Bigr) F^D_1(Q^2) 
\nonumber\\[2ex] F^{\null{iv}}_2(Q^2) &=  N\frac{5.7824 +
0.3907(1+Q^2/0.1422)^{-1}}{(1+Q^2/0.5362)}
   F^\rho_2(Q^2)\nonumber\\
 &\qquad{}  + \kappa_{\rho'} \frac{g_{\rho'}}{f_{\rho'}}
\frac{m^2_{\rho'}}{m^2_{\rho'} + Q^2} F^\rho_2(Q^2)
+ \Bigl( \kappa_\nu -6.1731\,N -  \kappa_{\rho'}\frac{g_{\rho'}}{f_{\rho'}} \Bigr)
F^D_2(Q^2)
\label{e10}
\end{align}
For $N=1$ the numerical values in Eq.~10 are those of \Cite{MMD} and are 
similar to those of \Cite{HO}.  They are determined by pion form factor and
pion-nucleon p-wave phase shift input into the dispersion relation  \Cite{MMD}.  
Because this input has uncertainties and is truncated at high momentum transfer, 
we considered the effect of an overall normalization factor N (the same for 
$F^{\null{iv}}_1$ and $F^{\null{iv}}_2$).

Because of the dispersion relation $\rho$ meson term, these models are 
labeled by DR-GK.  Model DR-GK(3) has the hadronic form factors of Model GK(3) 
(Eq.~(\ref{e6})).  Model DR-GK(1) uses the hadronic form factors of Model GK(1)
(Eq.~(\ref{e8})).   Model DR-GK$'$(1) and DR-GK$'$(3) are like Models DR-GK(1) and
DR-GK(3) respectively  but use the $\tilde Q^2$ of Eq.~(\ref{e9}).

The best fit value of $N$ varied between 0.78 and 0.94 for these models, but 
$\chi^2$ decreased substantially only for model DR-GK(3).  Consequently we present 
the results for the other three models with $N=1$, only introducing the extra 
parameter for Model DR-GK(3), now called DRN-GK(3). 

\section{Data base and fitting procedure}\label{s:3}

	The data for $G_{M_p}$ is from \Cite{LA,RCW,RGA1,PEB2,FB,WB1,KMH,Berger}.  The 
$G_{E_p}$ data is that of \Cite{LA,RCW,FB,KMH,Berger,JJM}.

The data sources for  $G_{M_n}$  are 
\Cite{KMH,Anklin,Bruins,Gao,AL,PM,ASE,Ank2,WB2,WB3,XU}.  The $G_{E_n}$ data is 
derived from \Cite{KMH,AL,WB2,WB3}, \Cite{JG,CH,MO,IP,DR,TE,MM,J-W}.  Recent small
revisions  in the published values of \Cite{MO,DR,MM} are included \Cite{Walcher}. 
Quasi-elastic deuteron and ${}^3$He data has been included, but the elastic deuteron
data  has been omitted because of its great sensitivity to the deuteron wave function. Another datum
is the slope $d G_{En}/dQ^2$ ($Q^2=0$) $=0.0199\pm0.0003$~fm$^2$, as determined by
thermal neutron scattering~\Cite{SK}.

	The data set for the ratio $R_p$ includes not only \Cite{MKJ}, which measures 
the ratio directly in a polarization experiment, but also the data of \Cite{WB1}, 
which extracts the ratio from unpolarized data dominated by the magnetic 
scattering.

	There are 11 free parameters in each of the models; the three~$g_m/f_m$ and the 
three $\kappa_m$ for the $\rho$ or $\rho'$, $\omega$ and $\phi$ mesons,
$\Lambda_1$, 
$\Lambda_2$, $\Lambda_D$, $\Lambda_{\mathrm{QCD}}$ and $\mu_\phi$.  Model
DRN-GK(3) has a 12th  parameter, $N$.  They were fitted by minimizing the value of
$\chi^2$ for all the  data using a Mathematica program that incorporates the
Levenberg-Marquardt  method.

\section{Results}\label{s:4}

	Table~\ref{T1} presents the ``best fit" parameters  to the present data set for 
the above 7 models.  The parameters of \Cite{GK} as fitted to the data set used 
in that reference are included in parentheses for Models GK(1) and GK(3).  For 
all but two of the seven models ``best fit" implies, as usual, the lowest local 
minimum in the search over the parameters.  However for Models DRN-GK(3) and
DR-GK$'$(3) the minimum is associated with indefinitely increasing negative values 
of $\kappa_\rho'$.  But $\chi^2$ decreases negligibly ($<1$\%) after reaching
reasonable values  of  $\kappa_\rho'$ which we choose to represent those two
models.

	As the models are simplifications of the actual physical situation, it is 
not required that the fitted parameters correspond to the values expected of 
them from  measurement of independent observables.  However those models for 
which the parameters are near those expectations are most consistent with the
known physics.  Only  four of the models, GK(1), DRN-GK(3), DR-GK(1) and
DR-GK$'$(1) have
$\Lambda_{QCD}$ in  the range of 100--300~MeV consistent with high energy
experiment.  The value of $\kappa_\rho$ is only a free parameter for the three GK
models.  Its value is  reasonable for GK(3) and  GK$'$(3) but is much too large for
GK(1).  Therefore only the above three DR-GK models
are consistent with the expected values of both $\Lambda_{\mathrm{QCD}}$ and
$\kappa_\rho$ .   Unfortunately none of the models have the expected small
negative value of
$\kappa_\omega$.  This is probably indicative that at least one higher mass  
isoscalar meson is important to the form factor description (the
$\omega$ and
$\phi$ meson widths are too small to require a modification of the pole 
representation).  Rather than further complicating the models, the isoscalar 
pole terms are to be regarded as effectively representing the more complicated 
situation of including higher mass isoscalar vector meson exchanges.  The stability
and adequacy of the fits is an indication that the  form factors with more poles
would be similar to those already obtained.

	In Table~\ref{T2} the values of $\chi^2$ are listed for all the models and the 
contribution from each of the five form factor classes of measurement (see 
beginning of Section~III) are detailed.  For the GK type models
$348.5 < \chi^2 < 352.8$ and for the DR-GK type $322.5 < \chi^2 < 327.1$.  Therefore
the  quality of the fit is essentially the same within a model type, but the models 
that use the $\rho$ meson contribution as determined by dispersion relations (and 
substitute the parameterized $\rho'$ pole contribution for the parameterized
$\rho$) are  significantly better fits to the data.  Within a model type there are
large  differences in the fitted parameters and important differences in the 
distribution of $\chi^2$ contributions among the different form factors, in spite 
of the small variation of the total values of $\chi^2$.  But the $\chi^2$ contributions
differ little for the  three models, DRN-GK(3), DR-GK(1) and DR-GK$'$(1), favored by
their physical values of $\Lambda_{QCD}$ and the dispersion  representation of the
$\rho$ meson contribution.  We also note that while Model  DRN-GK(3) has the
smallest value of $\chi^2$, it is the only one incorporating a  12th parameter, the
$\rho$ normalization $N$.  With $N=1$ the best $\chi^2$ for this  model is 375.1.  
By contrast the value of $\chi^2$  only decreases by 3 if $N$ is  allowed to vary in
the other three DR-GK type models.

	We note that with the parameters of the original GK fit \Cite{GK} the value of
$\chi^2$ with  respect to the present data set is 2.4 times larger than the best fit
value for  GK(1) and 3.0 times larger for GK(3).  Therefore the data accumulated
since 1992  has made an important difference.  We also note that while the best fit
values  of $\chi^2$ are about twice the number of degrees of freedom, this excess is 
mostly due to clear inconsistencies in the data sets, most particularly for $G_{Mn}$ 
at $Q^2 < 0.8~\mbox{GeV}/c^2$.  The displacement of nearby data points well beyond
their  given error bars is evident in the figures below.  \Cite{MMD} quotes a 
$\chi^2/$datum of~1.1.  As their fit is similar to those given here, this 
disparity may be due not only to the data accumulated since 1995 but also to the 
compactification in their case of many low momentum transfer points into slopes 
of the form factors at the origin.  Indeed, for the DRN-GK(3) model two points, at 
0.33 and 0.81 GeV$^2/c^2$, deviate in opposite directions for $G_{En}$ contributing
43.6  to a $\chi^2$ of 63.9.  For the same model 8 points, ranging from 0.24 to 0.81 
GeV$^2/c^2$, deviate from $G_{Mn}$ with both signs and contribute 89.6 to a $\chi^2$
of  120.1.  The results are similar for the other models.  This makes it clear that 
without the severe fluctuations of the experimental values outside their stated 
errors the fits presented here have achieved a value of $\chi^2$ close to the 
number of degrees of freedom.

	The following Figs.~\ref{ELFig1}--\ref{ELFig5} display the results for $G_{Mp}$, $G_{Ep}$,
$R_p$,  $G_{Mn}$,  and $G_{En}$, in that order.  $G_{Ep}$ and $G_{En}$ are normalized
to the dipole form factor  $G_d=(1+Q^2/0.71)^{(-2)}$.  $G_{Mp}$ ($G_{Mn}$) are
normalized to the product of $G_d$ and 
$\mu_p$ ($\mu_n$).  The Models GK(3), GK(1) and GK$'$(1) as fitted to the present
data  are compared in Figs. 1(a)--5(a), while in Figs. 1(b)--5(b) the same GK(3) and 
GK(1) are compared to those models with the parameters originally obtained in 
\Cite{GK}.  Figs.~1(c)--5(c) compare the results of Models DRN-GK(3), DR-GK(1), 
DR-GK$'$(1) and DR-GK$'$(3) with the data.  Figs.~6(a)--(c) show how all seven
models  extrapolate up to $Q^2 = 8$~GeV$^2/c^2$ for $R_p$ and the neutron form
factors, for  which data is now restricted to $Q^2 < 4$~GeV$^2/c^2$.  For those
observables we may expect  data at higher momentum transfers in the near future.

Figs.~7(a)--(d) show $G_{Mp}$, $G_{Ep}$, $G_{Mn}$,  and $G_{En}$ respectively for
the three favored models in the reduced range $Q^2 < 2$~GeV$^2/c^2$ where the
data was very crowded in the previous figures.  This gives a better view of the
model differences and the scatter of experimental points at low  $Q^2$.

	For $G_{Mp}$ (Fig.~\ref{ELFig1}) all the models agree closely over the very large 
momentum transfer range up to 31 GeV$/c^2$.  As shown in Fig.~1(b) even the 
substantial change in the fitted parameters from those of \Cite{GK}, which cause 
major differences in other form factors, make only a moderate difference here. 
But it should be noted that GK(3)-original is substantially lower at the peak 
near 2.5 GeV$/c^2$ than all the other models.  In this same momentum transfer 
region there is also a dichotomy in the experimental points.  There are some 
that peak near 1.06 (\Cite{LA,WB1}) and others that peak near 1.03 
(\Cite{RCW,RGA1}).  The fits of all the present models favor the higher values.  We
also note (Fig.~7(a)) that the model DRN-GK(3) is slightly favored by the data for
$Q^2 < 0.5$~GeV$^2/c^2$.

	The three GK type models are very close for $G_{Ep}$, while the four DR-GK type 
models have more spread at $Q^2 > 5$~GeV$/c^2$ (but still insignificant compared to 
experimental errors in that region).  For $G_{Ep}$, GK(3)-original is remarkable for
its  divergence from the present fits and GK(1)-original.  This may be due to an 
emphasis in~\Cite{GK} on fitting the data of \Cite{RCW} at $Q^2$ of 2.003, 2.497, and 3.007 
GeV$^2/c^2$, which were published shortly before~\Cite{GK} and in part motivated the 
variation of the GK(3) parameterization from that of GK(1).  This data is 
substantially higher in value than other data sets in the same range of momentum 
transfer \Cite{LA,WB1,Berger} that were published earlier.  It is to be noted that
at very low $Q^2 < 0.3$~GeV$^2/c^2$ the data \Cite{FB} is systematically lower
than the predictions of all the models (only the three favored models are in this
expanded figure) and the trend of the data for $Q^2 > 0.3$~GeV$^2/c^2$.  If this old
(1975) data is correct it implies that the models of charge distribution are
inadequate at ranges beyond 0.5 fm.  

	The presented $R_p$ data in Figs.~3 is independent of the $G_{Mp}$ and $G_{Ep}$
data  of Figs.~1 and~2.  The experiment of \Cite{WB1} and the polarization data of 
\Cite{MKJ}, which measure this ratio directly, are included only in these 
figures.  It is noticeable in comparing Figs.~3(a) and 3(c), and evident from 
the $\chi^2$ values (Table~\ref{T2}), that the DR-GK model fits are somewhat better
than  those of the GK models.  Fig.~3(b) shows, as in the case of $G_{Ep}$ discussed
above, that the  GK(3)-original model was too constrained by one particular set of
data.  The  extrapolation of this fit to 8 GeV$^2/c^2$ for $R_p$, Fig.~6(a), shows that
this  observable may be able to discriminate between the models at the higher
$Q^2$ if  the experimental errors do not increase at the higher momentum
transfers.  Even  the models preferred for their fit and physical parameters,
DRN-GK(3), DR-GK(1) and DR-GK$'$(1),  differ by as much as 0.1 at 8 GeV$^2/c^2$.

	Examining Figs.~4(a) and 4(c) one notes that while the overall fit to the 
$G_{Mn}$ data is about the same for all models, the GK models converge near
$Q^2=4$~GeV$^2/c^2$ while the DR-GK models diverge there.  Extrapolating to
8~GeV$^2/c^2$,  Fig.~6(b), the parameter favored models differ by almost 0.2, an
accuracy that  may be more achievable experimentally than the above mentioned
split for $R_p$.  Fig.~7(c) highlights the inconsistency of the $G_{Mn}$ experiments
for $Q^2 < 1.0$~GeV$^2/c^2$

	For $G_{En}$ Figs.~5(a) and 5(c) show that the improved fit of the DR-GK over the 
GK type models is most evident at the higher $Q^2$.  Extrapolating to
8~GeV$^2/c^2$, Fig.~6(c), there is little difference among the three parameter
favored  models.  Again, the expanded Fig.~7(d) shows that there is considerable
ambiguity in the low  $Q^2$ data.  In that figure we have added a new data point
at $0.495$~GeV$^2$ /Cite{HZ} which became available too late to be included in the
minimization of $\chi^2$.  As the value is within 1 st. dev. of the model curves, its
inclusion in the minimization would have made a negligible difference to the
parameter fit.

\section{Conclusions}\label{s:5}
	Moderately good fits to the nucleon electromagnetic form factor data are achieved for 
seven variations and extensions of the Gari-Kr\"umpelmann type model 
\Cite{GK} which preserves VMD at low momentum transfers and PQCD behavior at 
high momentum transfers.  The models all have simple analytic forms which are 
easily incorporated into few-nucleon form-factor predictions.

	 The four models which include the width of the $\rho$ meson, by use of 
dispersion relations, and the $\rho'$(1450) meson pole are a substantially better 
fit to the data than the $\rho$, $\omega$, and $\phi$ meson pole only GK models. 
The  fitted parameters of three of the four, DRN-GK(3), DR-GK(1), and DR-GK$'$(1),
have  values most compatible with independent evaluations.  For these three models
the  predictions for the nucleon electromagnetic form factors are not only 
quantitatively similar over the range of the present experimental data, but 
differ little when $R_p$, $G_{Mn}$, and $G_{En}$ are extrapolated to 8~GeV$^2/c^2$.  
Consequently only small differences due to the nucleon form factors are expected 
in predictions of deuteron emff (for which there is already $A(Q^2)$ data up to $Q^2 =
6$GeV$^2/c^2$)  and other few-nucleon electromagnetic form factors.   This will
eliminate a major ambiguity in the extraction of information about the 
few-nucleon wave functions and meson-exchange current effects.  Precise data in 
the $Q^2 = 4$--8 GeV$^2/c^2$ range may eventually further narrow the uncertainty.

\section*{acknowledgments}

	The author is grateful to Manfred Gari for discussion of the development 
of the model and to Haiyan Gao for information about new form factor data and 
re-evaluations of past data.

\begin{figure}[p]
$$
\BoxedEPSF{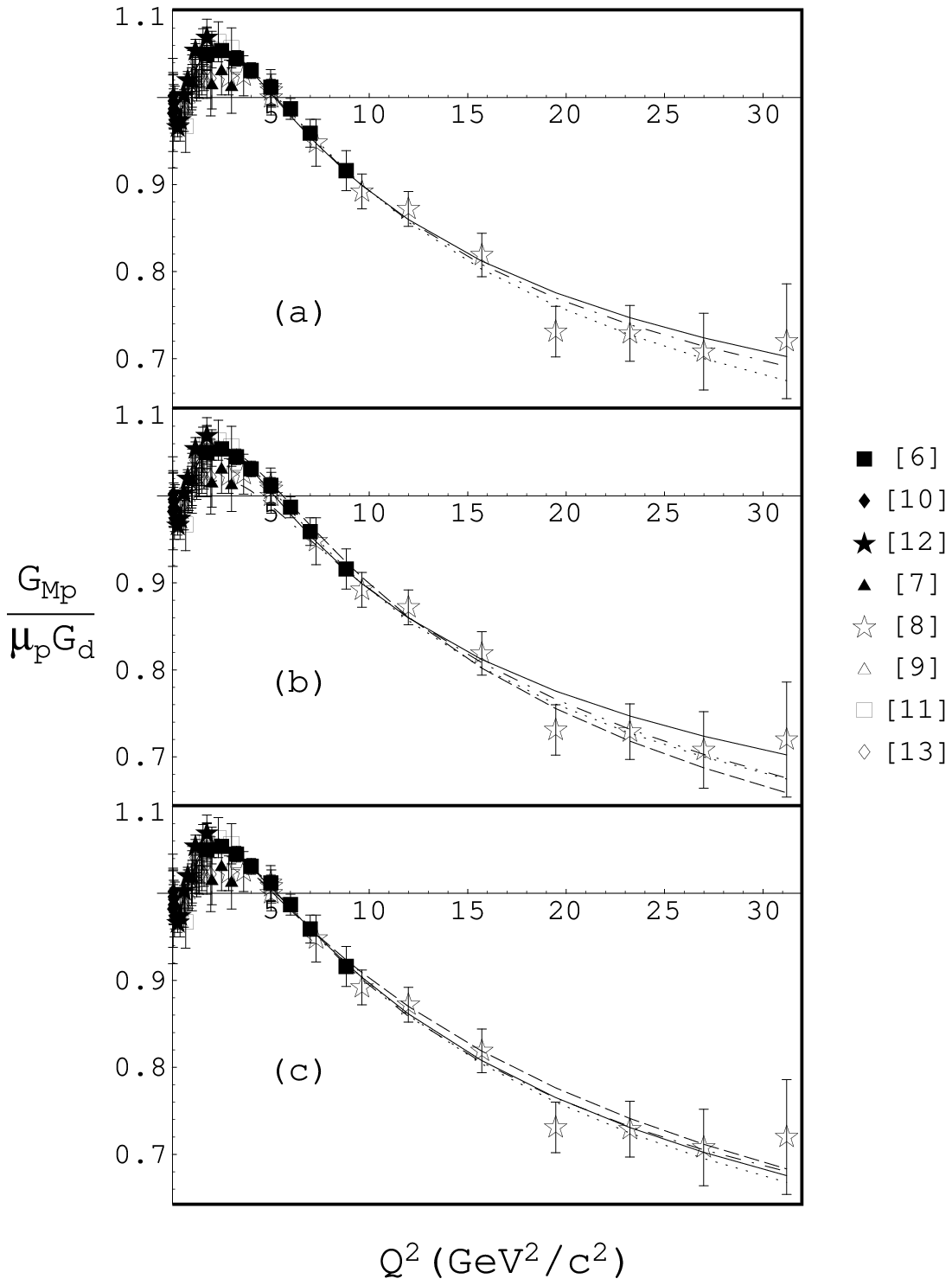 scaled 1200}
$$
\caption{$G_{Mp}$ normalized to $\mu_p G_d$. (a) Comparison of the models 
GK(3) [solid], GK(1) [dotted] and GK$'$(1) [dash-dotted] with the data.  (b) 
Comparison of GK(3) [solid] and GK(1) [dotted] with the same models and the 
parameters of \protect\Cite{GK}, GK(3)-original [dash-dotted] and GK(1)-original [dashed].  
(c) Comparison of models DRN-GK(3) [solid], DR-GK(1) [dotted], DR-GK$'$(1) [dash-dotted] and
DR-GK$'$(3) [dashed] with the data.}
\label{ELFig1}
\end{figure}

\begin{figure}[p]
$$
\BoxedEPSF{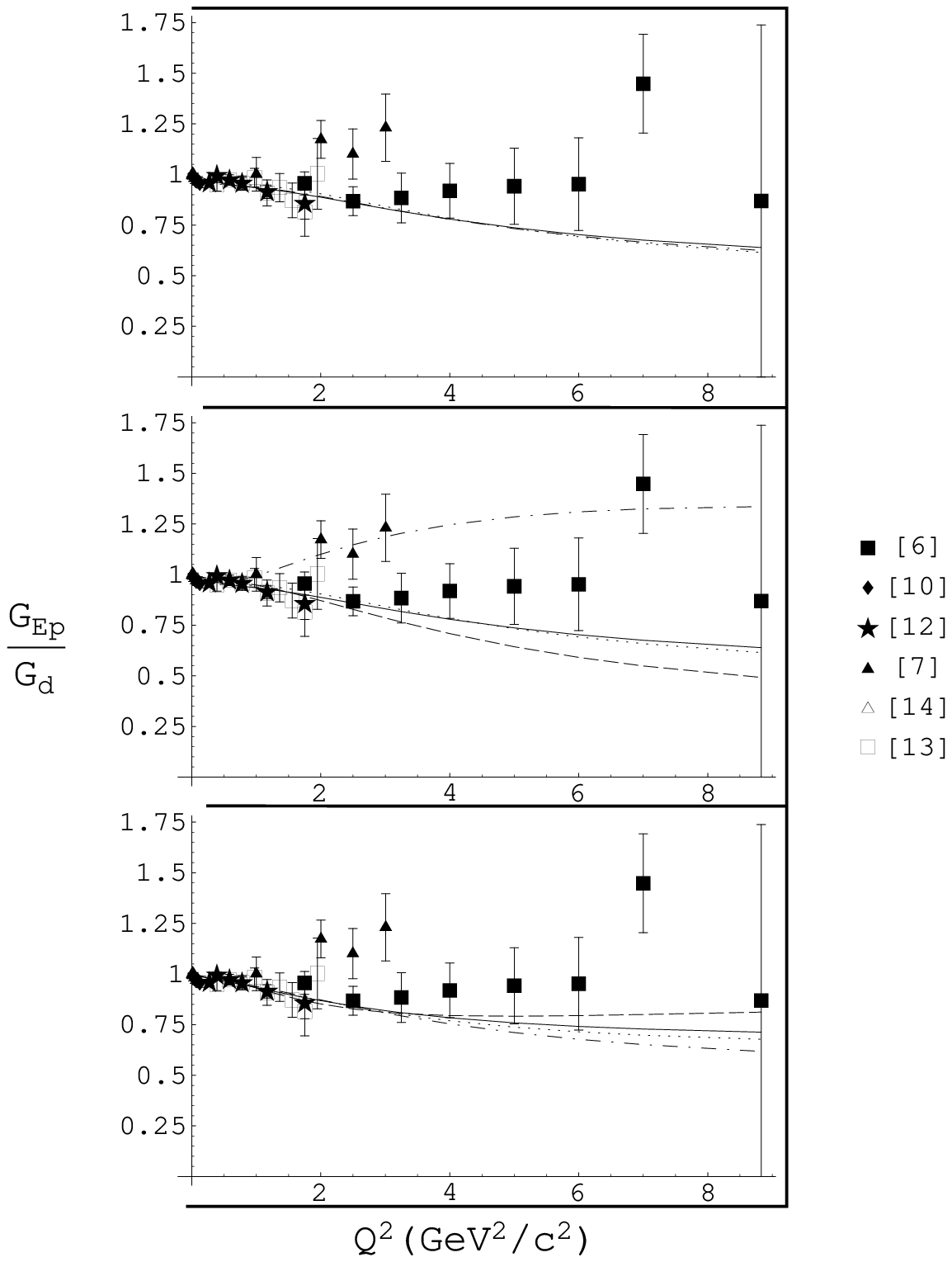 scaled 1200}
$$
\caption{$G_{Ep}$ normalized to $G_d$.  (a) Comparison of the models 
GK(3)[solid], GK(1)[dotted] and GK$'$(1)[dash-dotted] with the data.  (b) 
Comparison of GK(3)[solid] and GK(1)[dotted] with the same models and the 
parameters of \protect\Cite{GK}, GK(3)-original[dash-dotted] and GK(1)-original[dashed].  
(c) Comparison of models DRN-GK(3)[solid], DR-GK(1)[dotted], DR-GK$'$(1)[dash-dotted] and
DR-GK$'$(3)[dashed] with the data.}
\label{ELFig2}
\end{figure}

\begin{figure}[p] 
$$
\BoxedEPSF{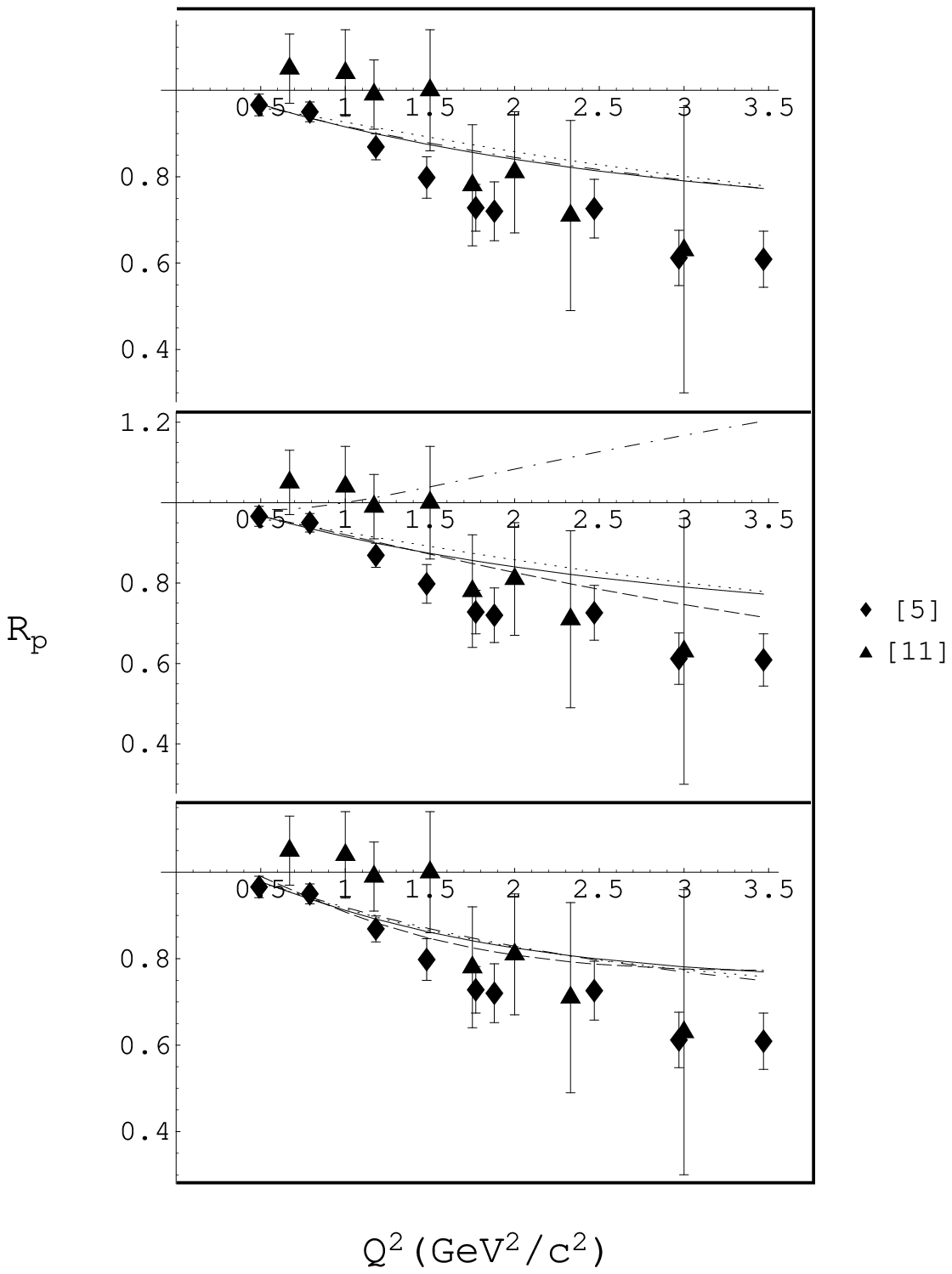 scaled 1200}
$$
\caption{$R_p$, the ratio $\mu_p G_{Ep}/G_{Mp}$.  (a) Comparison of the models 
GK(3) [solid], GK(1) [dotted] and GK$'$(1) [dash-dotted] with the data.  (b) 
Comparison of GK(3) [solid] and GK(1) [dotted] with the same models and the 
parameters of \protect\Cite{GK}, GK(3)-original [dash-dotted] and GK(1)-original [dashed].  
(c) Comparison of models DRN-GK(3) [solid], DR-GK(1) [dotted], DR-GK$'$(1) [dash-
dotted] and DR-GK$'$(3) [dashed] with the data.}
\label{ELFig3}
\end{figure}

\begin{figure}[p] 
$$
\BoxedEPSF{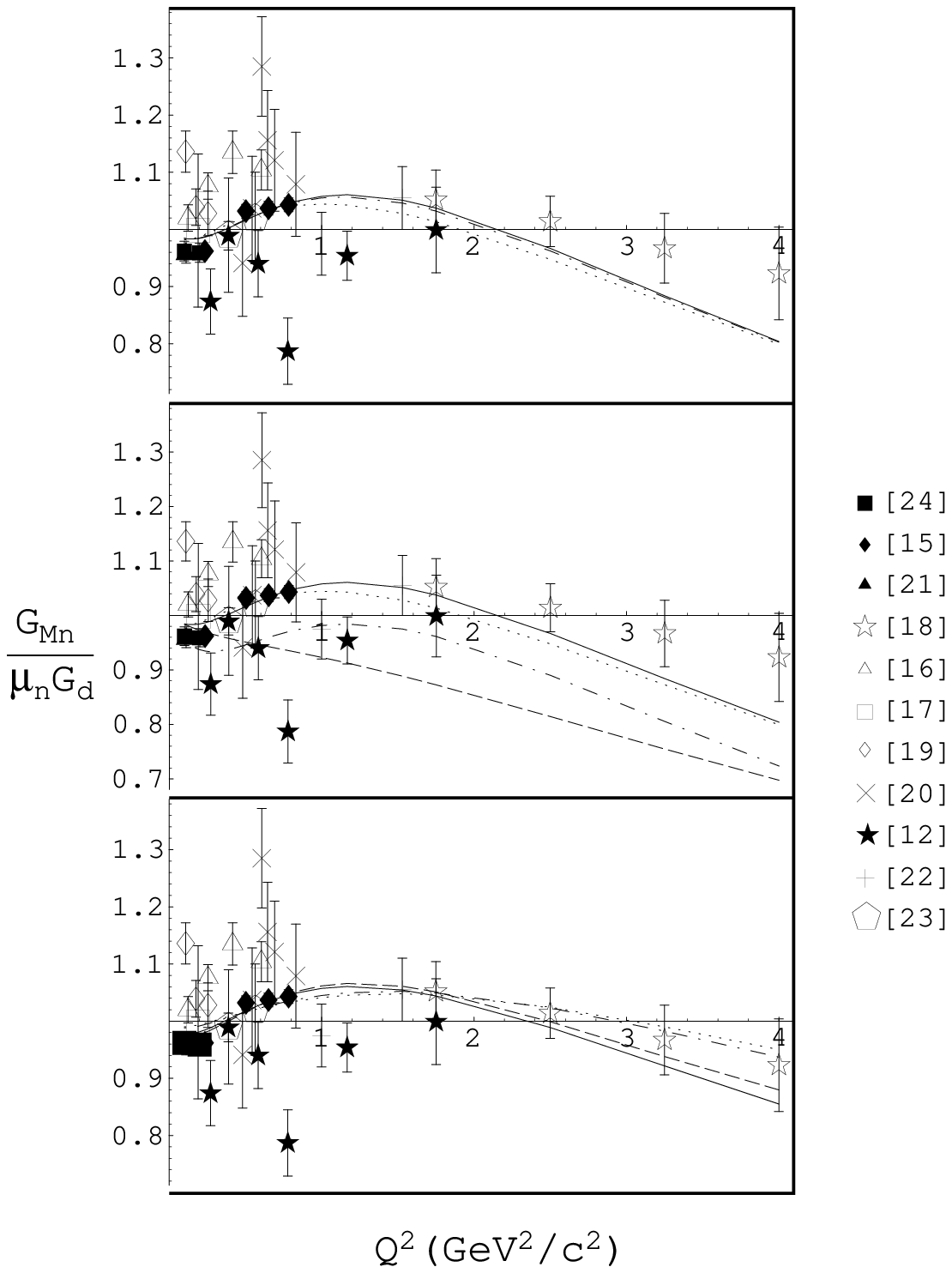 scaled 1200}
$$
\caption{$G_{Mn}$ normalized to $\mu_n G_{d}$.  (a) Comparison of the models 
GK(3) [solid], GK(1) [dotted] and GK$'$(1) [dash-dotted] with the data.  (b) 
Comparison of GK(3) [solid] and GK(1) [dotted] with the same models and the 
parameters of \protect\Cite{GK}, GK(3)-original [dash-dotted] and GK(1)-original [dashed].  
(c) Comparison of models DRN-GK(3) [solid], DR-GK(1) [dotted], DR-GK$'$(1) [dash-dotted] and
DR-GK$'$(3) [dashed] with the data.}
\label{ELFig4}
\end{figure}

\begin{figure}[p] 
$$
\BoxedEPSF{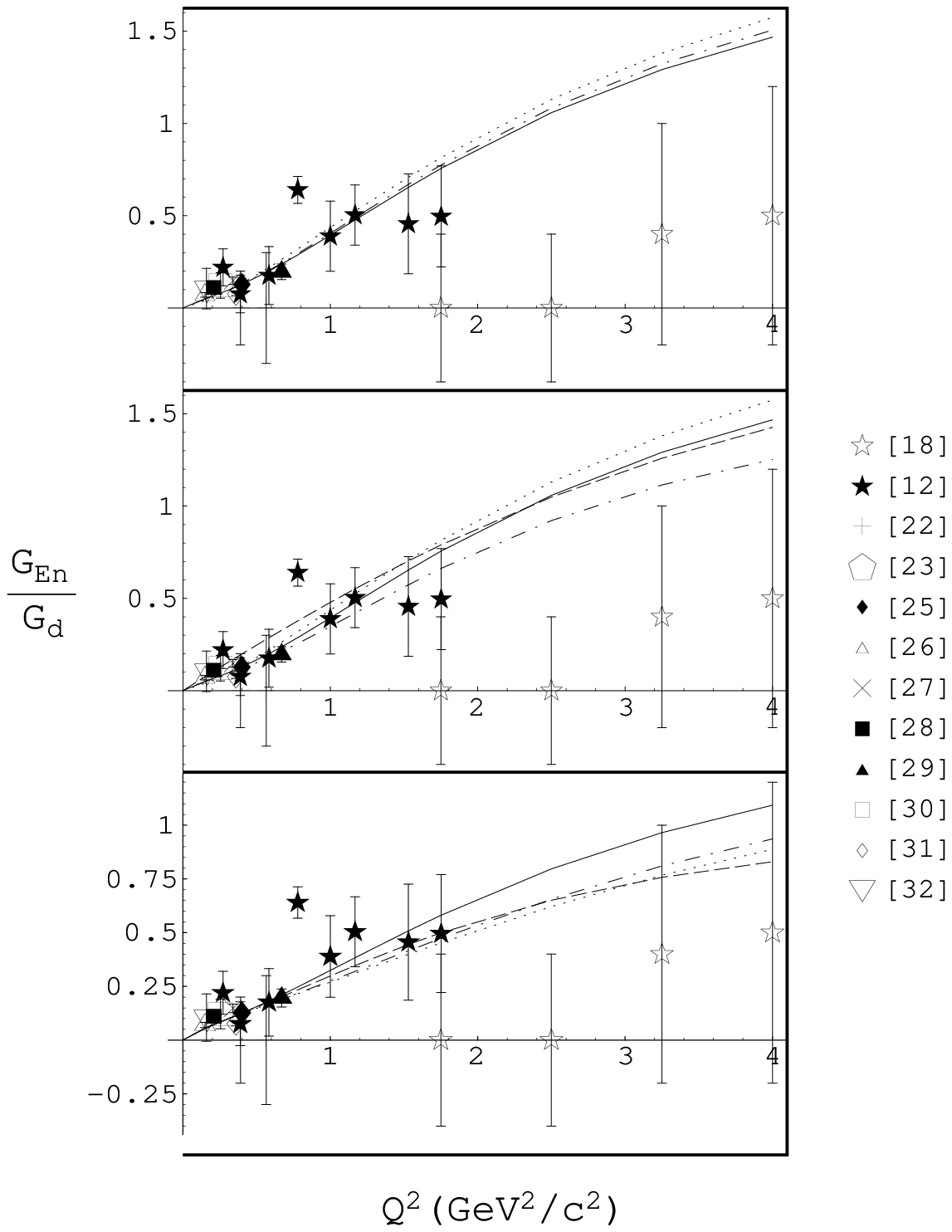 scaled 1200}
$$
\caption{$G_{En}$ normalized to $G_{d}$.  (a) Comparison of the models 
GK(3) [solid], GK(1) [dotted] and GK$'$(1) [dash-dotted] with the data.  (b) 
Comparison of GK(3) [solid] and GK(1) [dotted] with the same models and the 
parameters of \protect\Cite{GK}, GK(3)-original [dash-dotted] and GK(1)-original [dashed].  
(c) Comparison of models DRN-GK(3) [solid], DR-GK(1) [dotted], DR-GK$'$(1) [dash-dotted] and
DR-GK$'$(3) [dashed] with the data.}
\label{ELFig5}
\end{figure}

\begin{figure}[p]
$$
\BoxedEPSF{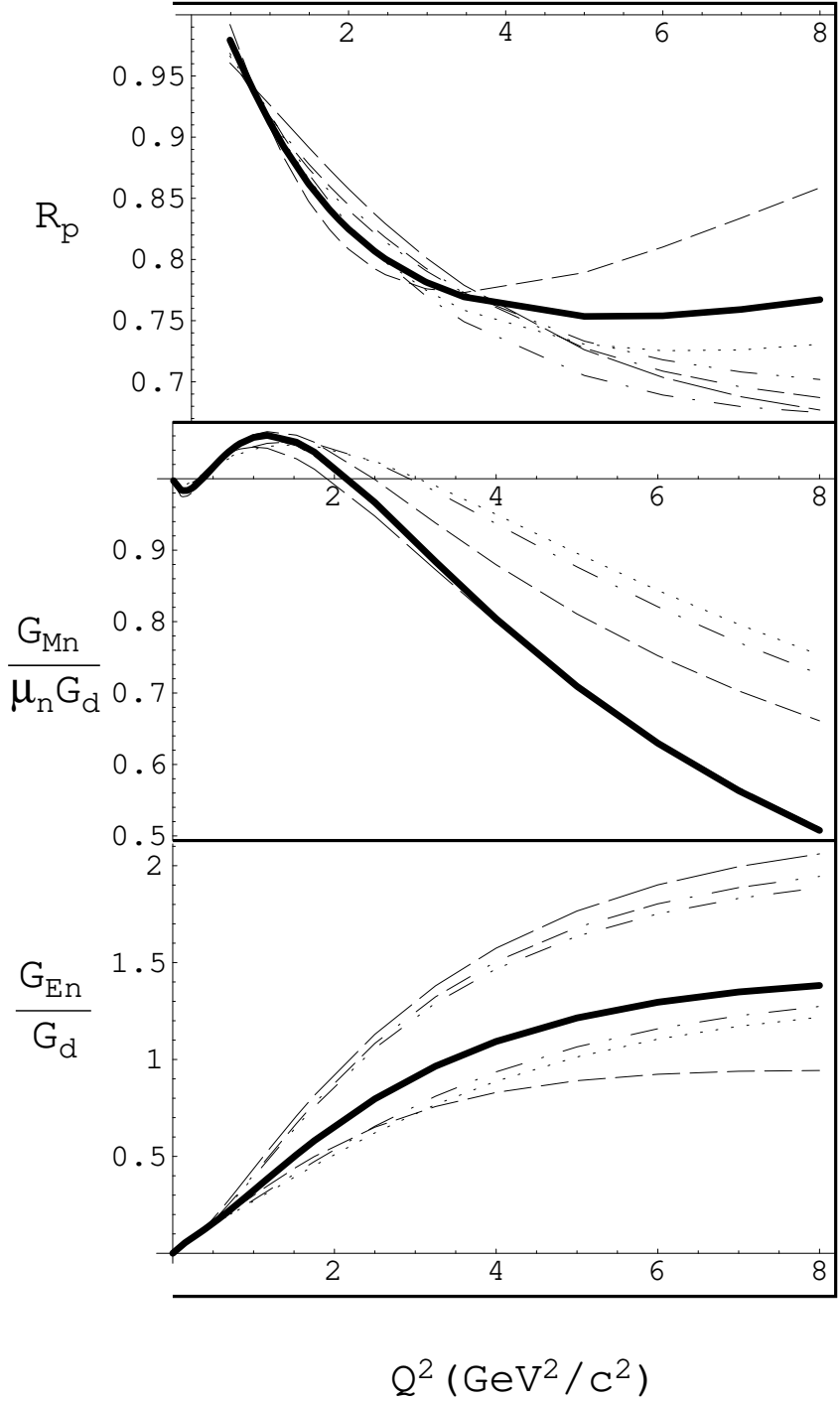 scaled 1200}
$$
\caption{Extrapolation to $Q^2 = 8$~GeV$^2/c^2$.  Comparison of the models 
DRN-GK(3) [solid], DR-GK(1) [dotted], DR-GK$'$(1) [dash-dotted], DR-GK$'$(3) [dashed], 
GK(3) [dash-double dotted], GK(1) [long dashes] and GK$'$(1) [double dash-dotted].  
(a) $R_p$, the ratio $\mu_p G_{Ep}/G_{Mp}$.  (b) $G_{Mn}$ normalized to $\mu_n G_{d}$.   
(c) $G_{En}$
normalized to $G_{d}$.}
\label{ELFig6}
\end{figure}

\begin{figure}[p]
$$
\BoxedEPSF{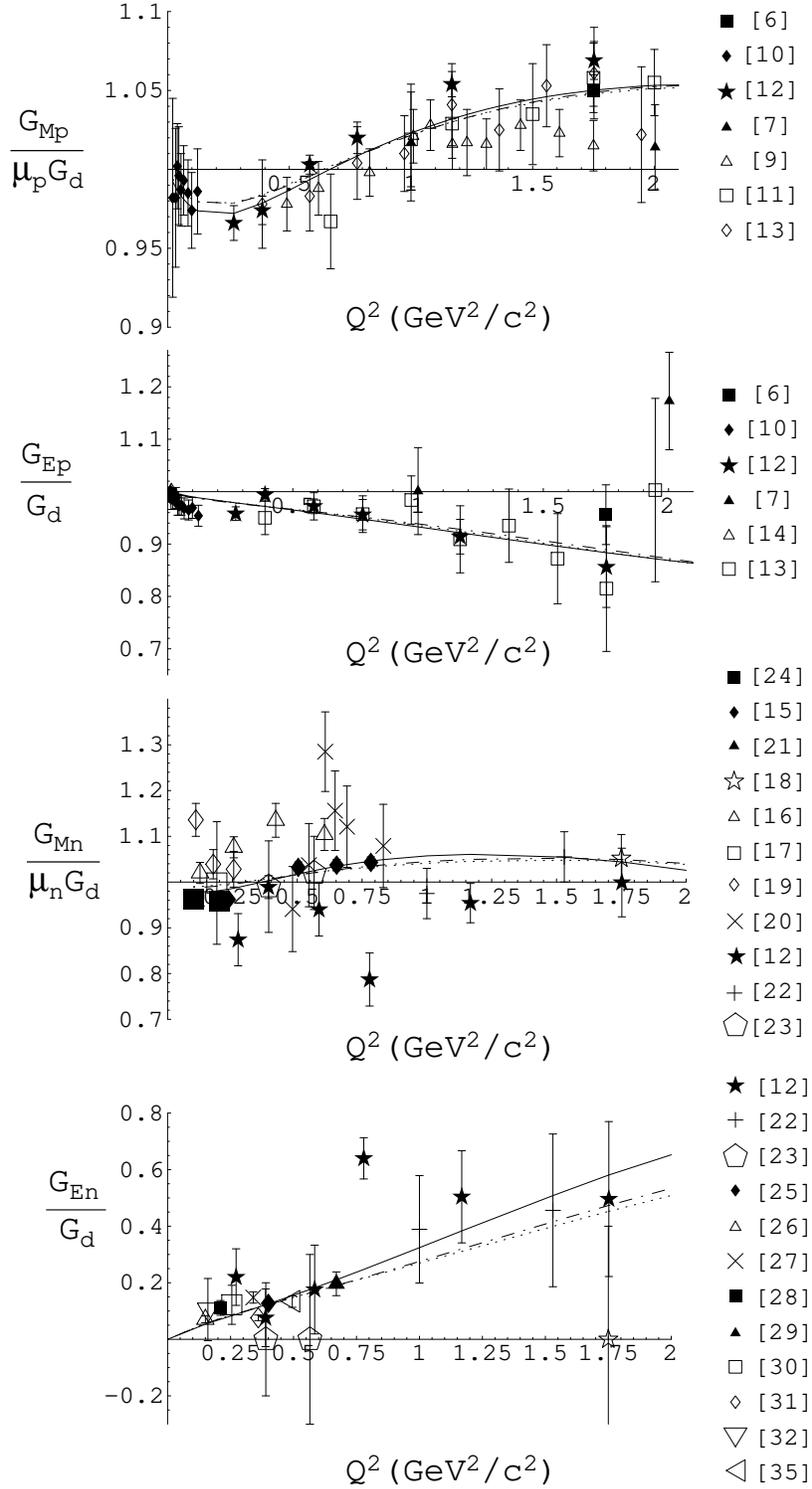 scaled 945}
 $$
 \caption{Expanded interval $Q^2 < 2$GeV$^2/c^2$.  Comparison of the
favored models  DRN-GK(3)  [solid], DR-GK(1) [dotted], DR-GK$'$(1)  [dash-dotted].
(a) $G_{Mp}$ normalized to $\mu_p G_{d}$.   (b) $G_{Ep}$ normalized to $G_{d}$.  
(c) $G_{Mn}$ normalized to $\mu_n G_{d}$.   (d) 
$G_{En}$ normalized to $G_{d}$.  The point at 0.495 GeV$^2/c^2$ \protect\Cite{HZ} was
added after optimization of the model parameters and appears only in this figure.}
 \label{ELFIG. 7}
 \end{figure}

 \begin{table}[p] 
 \caption{Model parameters.  Common to all models are $\kappa_v=3.706$, 
$\kappa_s=-0.12$, $m_\rho=0.776$ GeV, $m_\omega=0.784$ GeV, $m_\phi=1.019$
GeV and  $m_{\rho'}= 1.45$ GeV.  Parentheses contain the values of
\protect\Cite{GK}.}
 \label{T1}
 \begin{tabular}{c|ccccccc} 
Parameters & & & & Models & & & \\
\hline
 &  GK(3) & GK(1) & GK$'$(1) & DRN-GK(3) & DR-GK$'$(3) & DR-GK(1) & DR-GK$'$(1)
\\ (a) $g_{\rho^{(\prime)}}/f_{\rho^{(\prime)}}$  & 0.4466 & 0.0514 & 0.3223 &
0.1013 & 0.0808  & 0.0625  & 0.0636 \\ & (0.5688) & (0.377) & & & & & \\
(a) $\kappa_{\rho^{(\prime)}}$  & 4.3472 & 23.533  & 4.982 & $-15.870$ &
$-17.993$ & 0.9397 & $-0.4175$ \\ & (3.642) & (6.62) & & & & & \\ 
$g_\omega/f_\omega$ & 0.4713 & 0.0588 &0.3440 & 0.6604 & 0.8038 & 0.8029 &
0.7918 
\\ & (0.5774) & (0.411) & & & & & \\ $\kappa_\omega$ & 21.762 & 18.934 &
40.661 &  8.847 & 4.0526 & 5.5225 & 5.1109 \\ & (0.4775) & (0.163) & & & & & \\ 
$g_\phi/f_\phi$ & $-0.8461$ & $-0.5283$ & $-0.9315$ & $-0.4054$ & $-0.2336$ &
$-0.3070$ & $-0.3011$ \\ & $(-0.666)$ & (0.0) & & & & & \\  $\kappa_\phi$ &
11.849 & 1.2236 &  14.6805 & 13.6415 & 13.5963 & 14.4123 & 13.4385 \\ &
$(-0.2378)$ & (0.0) & & & & & 
\\ $\mu_\phi$ & 1.1498 & 1.1670 & 1.1411 & 1.127 & 1.1218 & 1.2379 & 1.1915 \\
&  (0.33) & (--) & & & & & \\ $\Lambda_1$ & 0.9006 & 0.5902 & 0.8956 & 0.89361
&  0.9295 & 0.9916 & 0.9660 \\ & (0.823) & (0.795) & & & & & \\ $\Lambda_D$ &
1.7038 &  0.7273 & 1.7038 & 1.0454 & 1.2207 & 1.2589 &1.3406 \\ & (1.24) &
(0.795) & & & &  & \\ $\Lambda_2$ & 1.1336 & 1.9368 & 0.9551 & 2.1614 & 3.9736
& 2.1327 & 2.1382 \\  & (1.95) & (2.270) & & & & & \\ $\Lambda_{\mathrm{QCD}}$
& 0.0312 & 0.1377 & 0.0604 & 0.2452  & 0.4394 & 0.1377 & 0.1163 \\ & (0.31) &
(0.29) & & & & & \\ $N$ & -- & -- & -- &  0.7838 & 1.0 (b) & 1.0 (b) & 1.0 (b) \\ 
 \end{tabular}
$^{\mathrm{(a)}}$ 
$\rho^{(\prime)}$ signifies the $\rho$ meson  for the GK models and the
$\rho'(1450)$ meson for the DR-GK models.  $^{\mathrm{(b)}}$  not varied
 \end{table}

 \begin{table}[p] 
 \caption{Contributions to the standard deviation, $\chi^2$, from each data type for
each of the models. The number of data points, $n$, is listed for each data type.
Parentheses contain results of ref.~\protect\Cite{GK} parameters.}
 \label{T2}
 \begin{tabular}{r|c|rrrrrrr} 
Data &&& & & Models & & & \\[-0.5ex]
type & $n$\ \null &  \llap{G}K(3\rlap{)} & \llap{G}K(1\rlap{)} & \llap{G}K$'$(1\rlap{)} &
\llap{DR}N-GK(3\rlap{)} & \llap{DR}-GK$'$(3\rlap{)} & \llap{DR}-GK(1\rlap{)} &
\llap{DR}-GK$'$(1)\\
\hline\hline 
$G_{Mp}$ &\llap{6}8 & 47.6 & 45.4 & 45.6 & 42.3
& 46.7  & 42.9  & 43.3\ \null \\
&& (206.8\rlap{)} & (71.9\rlap{)} && && & \\
$G_{Ep}$ & \llap{4}8  & 72.7 & 65.2  & 71.8 & 65.8 & 68.0
& 65.8 &67.2\ \null \\
&& (97.1\rlap{)} & (76.2\rlap{)} && && & \\
$G_{Mn}$ &\llap{3}5 &
124.8 & 123.9 & 124.3 & 120.1 & 121.0 & 123.8 & 122.4\ \null \\
&& (344.1\rlap{)} & (393.9\rlap{)} && && & \\
$G_{En}$ &\llap{2}3 & 69.4 & 76.6 &
70.5 &  63.9 & 62.8 & 65.1 & 64.8\ \null \\
&& (69.7\rlap{)} & (217.5\rlap{)} && && & \\ 
$R_p$ &\llap{1}7 & 35.0 & 41.6 & 36.4 & 30.5 & 27.8 &
29.4 & 29.0\ \null \\
&& $(323.5\rlap{)}$ & (25.2\rlap{)} && && & \\
\hline 
Total &\llap{19}1  &
349.5 & 352.7 &  348.6 & 322.6 & 326.3 & 327.0 & 326.7\ \null \\
&&
\llap{(1},041.2\rlap{)} & (784.7\rlap{)} && && & 
 \end{tabular}

 \end{table}  

\end{document}